\documentclass[fleqn,twoside]{article} \usepackage{espcrc2}

\usepackage{graphicx}
\usepackage[figuresright]{rotating}

\newcommand{\text}[1]{\mbox{\tiny #1}}

\newcommand{\AmS}{{\protect\the\textfont2
  A\kern-.1667em\lower.5ex\hbox{M}\kern-.125emS}}

\title{Hadronic physics with domain-wall valence and improved
  staggered sea quarks}

\author{LHP Collaboration{\thanks{Supported by DOE contracts
          DE-FC02-94ER40818, DE-FG02-91ER40676, and DE-AC05-84ER40150.
          W.S.~is an Alexander von Humboldt fellow. Computations were
          performed on Intel Pentium IV Xeon clusters at Jefferson
          Laboratory, and at ORNL using time awarded under the
          auspices of the DOE's SciDAC initiative. We used MILC gauge
          field configurations}}:\\ D.B.~Renner\address[MIT]{Center
          for Theoretical Physics, Massachusetts Institute of
          Technology, Cambridge, MA 02139, USA}\thanks{Current
          address: Department of Physics, University of Arizona, 1118
          E 4th Street, Tucson, AZ 85721, USA},
          W.~Schroers\addressmark[MIT],
          R.~Edwards\address[JLAB]{Thomas Jefferson National
          Accelerator Facility, Newport News, VA 23606, USA},
          G.T.~Fleming\addressmark[JLAB],
          Ph.~H{\"a}gler\address{Department of Physics and Astronomy,
          Vrije Universiteit, Amsterdam, The Netherlands},
          J.W.~Negele\addressmark[MIT], K.~Orginos\addressmark[MIT],
          A.V.~Pochinski\addressmark[MIT], and
          D.~Richards\addressmark[JLAB]}
       
\begin{document}

\begin{abstract}
With the advent of chiral fermion formulations, the simulation of
light valence quarks has finally become realistic for numerical
simulations of lattice QCD\@. The simulation of light dynamical
quarks, however, remains one of the major challenges and is still an
obstacle to realistic simulations. We attempt to meet this challenge
using a hybrid combination of Asqtad sea quarks and domain-wall
valence quarks. Initial results for the proton form factor and the
nucleon axial coupling are presented.
\end{abstract}

\maketitle

\section{INTRODUCTION}
\label{sec:introduction}
The simulation of light dynamical quarks constitutes one of the major
challenges in contemporary lattice gauge theory research. The
computational obstacle has in the past been addressed by investigating
new algorithms, see, e.g.~\cite{Schroers:2001if} and references
therein.

Recently, a new approach has been attempted employing a so-called
``hybrid'' scheme where different types of sea and valence quarks are
used, see~\cite{Davies:2003ik}. It turned out that this calculation is
extremely successful in predicting spectroscopic mass splitting in
heavy quark physics correctly.

In this work we employ another type of hybrid calculation involving
improved Kogut-Susskind sea quarks (Asqtad
action~\cite{Orginos:1999cr}) and domain-wall valence
fermions~\cite{Kaplan:1992bt}. This scheme --- although breaking
unitarity at finite lattice spacing --- will still have the same
continuum limit as a fully dynamical calculation provided this limit
exists. The quark masses of sea and valence quarks have to be properly
tuned. The quantities we investigate are special cases of generalized
parton distributions (GPDs)~\cite{Muller:1998fv}. These have both
parton distributions and form factors as certain limits and have
already been studied on the lattice using conventional
schemes~\cite{Hagler:2003jd}.

This presentation is organized as follows: After discussing the tuning
of the parameters of the valence domain-wall action in
Section~\ref{sec:choice-parameters}, we present the main results in
Section~\ref{sec:hadr-struct-with}. Finally, we summarize our findings
and give an outlook for our ongoing research in
Section~\ref{sec:summary-outlook}.

\section{CHOICE OF PARAMETERS}
\label{sec:choice-parameters}
The two essential parameters we have to set are the size of the fifth
dimension, $L_5$, for the domain-wall fermions and the bare quark mass
parameter, $(am)_q^{\text{DWF}}$. The tuning of these two parameters
is discussed in this section. The resulting parameters and sample
sizes are summarized in Table~\ref{tab:tune-quarks}.
\begin{table*}[ht]
\caption{Parameters of our runs after the tuning of the DWF valence
  quark mass. The table shows the lattice volume, $\Omega$, the number
  of configurations in each sample, $\#$, the bare Asqtad masses for
  the sea and valence part, the bare domain-wall fermion masses, and
  the resulting ratio of pseudoscalar to vector meson masses.}
\label{tab:tune-quarks}
\begin{tabular}{@{}*{8}{l}}
\hline
$\Omega$ & $\#$ &
$(am)_q^{\text{Asqtad,sea}}$ & $(am)_q^{\text{Asqtad,val}}$ &
$(am)_q^{\text{DWF}}$ &
$m_{\pi}^{\text{Asqtad}}$ / MeV & 
$m_{\pi}^{\text{DWF}}$ / MeV &
$m_{\pi}/m_{\rho}$ \\ \hline
$20^3\times 32$ & $107$ & $0.050$ & $0.050$ & $0.0810$ & 
$774.8(0.3)$ & $775.8(2.1)$ & $0.687(6)$ \\
& $134$ & $0.030/0.050$ & $0.030$ & $0.0478$ &
$604.6(0.3)$ & $605.8(2.1)$ & $0.588(7)$ \\
& $56$  & $0.020/0.050$ & $0.020$ & $0.0313$ &
$498.0(0.3)$ & $502.1(3.7)$ & $0.530(11)$ \\
& $104$ & $0.010/0.050$ & $0.010$ & $0.0138$ &
$359.1(0.4)$ & $368.8(3.5)$ & $0.415(9)$ \\
$28^3\times 32$ & $138$ & $0.010/0.050$ & $0.010$ & $0.0138$ &
& $363.9(1.3)$ & $0.387(7)$ \\
\hline
\end{tabular}
\end{table*}

We used MILC configurations both from the NERSC archive and provided
directly by the collaboration. We then applied
HYP-smearing~\cite{Hasenfratz:2001hp} and bisected the lattice in the
time direction. Thus far, we have only considered the first
half-lattice. We have chosen the gauge field configurations separated
by $12$ trajectories. In these samples we did not find residual
autocorrelations.

\subsection{Setting $L_5$}
\label{sec:setting-l_5}
The goal in setting $L_5$ is to describe the physics adequately at
minimal computational cost. For finite values of $L_5$ there is a
residual explicit chiral symmetry breaking characterized by a residual
mass, $(am)_{\text{res}}$. We adopt the following definition
(see~\cite{Blum:2000kn} for details):
\begin{equation}
  \label{eq:ward-taka}
  \Delta^\mu {\cal A}_\mu^a = 2m_q J_q^a(x) + 2J_{5q}^a(x)\,,
\end{equation}
where
\begin{equation}
  \label{eq:mres-def}
  J_{5q}^a(x) \approx m_{\text{res}} J_5^a(x)\,,
\end{equation}
which holds up to ${\cal O}(a^2)$. We require $m_{\text{res}}$ to be
at least one order of magnitude smaller than $m_q$.

To explore their dependence, we have run simulations using two samples
of $25$ configurations with volume $\Omega=20^3\times 32$ from
Table~\ref{tab:tune-quarks}: three degenerate dynamical Asqtad quarks
with masses $(am)_q^{\text{Asqtad,sea}}=0.050$ (denoted as ``heavy'')
and two plus one quarks with masses
$(am)_q^{\text{Asqtad,sea}}=0.010/0.050$ (termed ``light'').

\begin{figure}[ht]
  \caption{Residual quark mass as a function of $L_5$ for the two
  samples (heavy and light) of $25$ configurations each.}
  \label{fig:mres}
  \vskip 7pt
  \includegraphics[scale=0.25,clip=true]{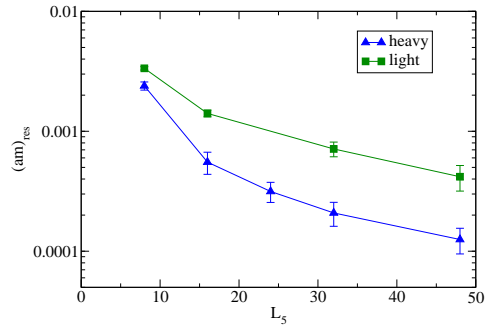}
\end{figure}

The resulting residual masses obtained from Eqs.~(\ref{eq:ward-taka})
and (\ref{eq:mres-def}) are plotted in Figure~\ref{fig:mres}. In the
light quark case, $L_5=16$ just fulfills our requirement, while in the
heavy quark case $L_5=16$ more than satisfies it.

\begin{figure}[ht]
  \caption{The pion mass with heavy and light quarks,
  $m_{\pi,\text{heavy}}$ and $m_{\pi,\text{light}}$, and the nucleon
  mass with heavy and light quarks, $m_{\text{N,heavy}}$ and
  $m_{\text{N,light}}$, as a function of $L_5$.}
  \label{fig:masses-L_5}
  \vskip 7pt
  \includegraphics[scale=0.25,clip=true]{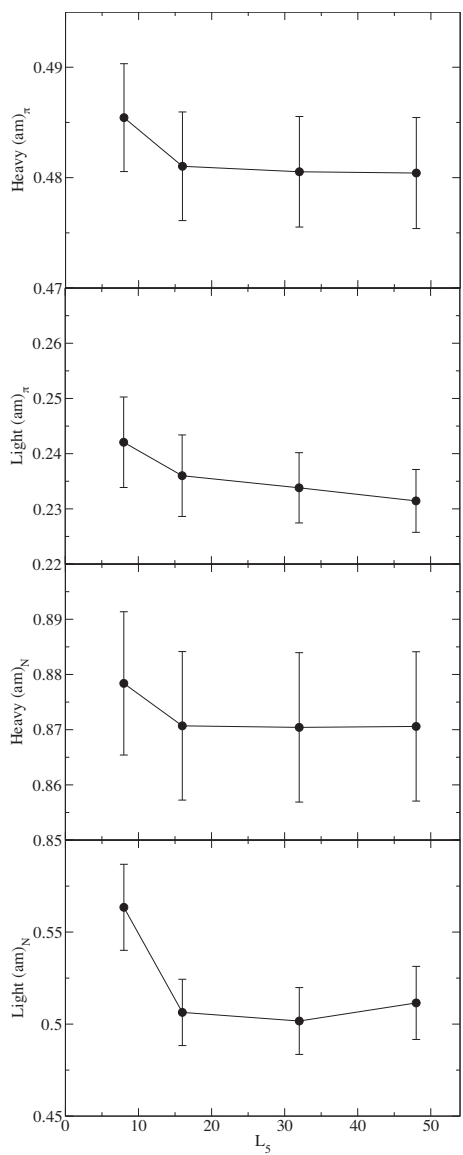}
\end{figure}

\begin{figure}[ht]
  \caption{The same quantities as in Figure~\ref{fig:masses-L_5}, but
    with a Jackknife analysis of the mass differences vs.~$L_5=16$.}
  \label{fig:diffm-L_5}
  \vskip 7pt
  \includegraphics[scale=0.25,clip=true]{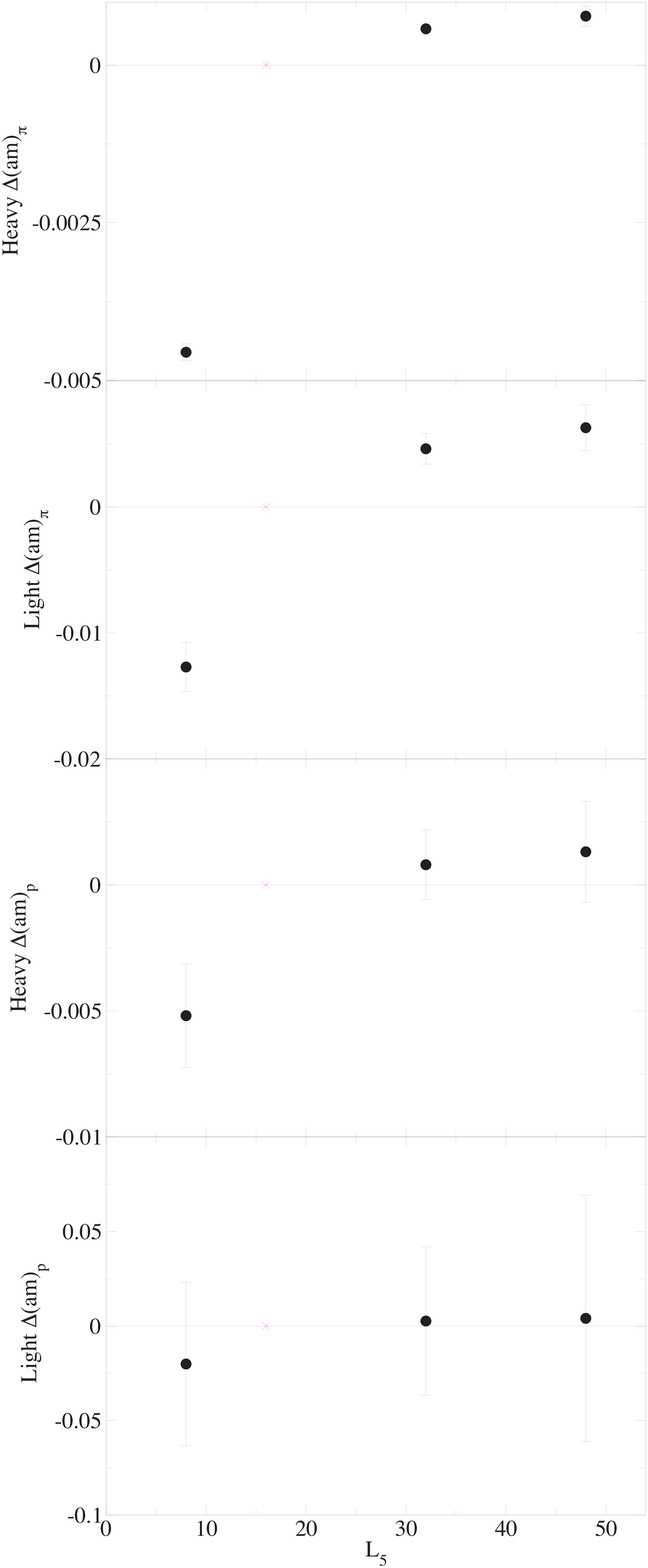}
\end{figure}

Furthermore, we require the absolute change in the masses of the pion
and the nucleon under changes of $L_5$ to be small; these are shown in
Figure~\ref{fig:masses-L_5} for the pion and nucleon masses. Note,
that the errors are not independent between different points of $L_5$,
since an identical sample of gauge field configurations has been
used. Therefore, we have performed a separate analysis by Jackknifing
the differences in the masses compared to $L_5=16$. The results are
shown in Figure~\ref{fig:diffm-L_5}.

In the case of heavy quarks, the influence of increasing $L_5$ beyond
$L_5=16$ is negligible in all observables. In the case of the light
quarks, the influence is at most a few percent when going beyond
$L_5=16$. Hence, we choose $L_5=16$ to be a good compromise between
accuracy and performance.
\clearpage

\subsection{Setting the quark mass}
\label{sec:setting-quark-mass}
As outlined in the introduction, a necessary condition for a hybrid
calculation to describe the physics of full QCD is that the masses of
sea and valence quarks are identical.

Hence, we have chosen to match the pion masses of two different
calculations: (i) full QCD calculations using three and two plus one
flavors of dynamical Asqtad sea fermions and Asqtad valence
fermions. These results have been taken
from~\cite{Bernard:2001av}. (ii) Our hybrid calculation with Asqtad
dynamical sea fermions and valence domain-wall fermions with $L_5=16$.

To make the quark masses coincide, we tuned the bare mass parameter in
the domain-wall valence fermion action so that the resulting
pseudoscalar meson mass coincides with the corresponding meson mass in
the Asqtad case. This condition is useful since the pseudoscalar meson
mass is particularly sensitive to the choice of the quark mass.

The resulting choices for the bare quark masses, $(am)_q$, for Asqtad
and domain-wall fermions are shown in Table~\ref{tab:tune-quarks}. It
is interesting that the difference between the bare quark masses for
Asqtad and DWF valence quarks is quite substantial. Since this
difference corresponds to the ratio of quark mass renormalization
constants, we observe that either one or both of these renormalization
constants deviates substantially from one.

Finally, we list the resulting nucleon masses --- which we need for
the calculation of the hadronic structure --- in
Table~\ref{tab:nucl-mass}. There is a perceptible difference at the
heaviest quark mass. The other masses are statistically
compatible. Furthermore, there is no visible finite-size effect at the
lightest quark mass. Thus, we only observe the expected ${\cal
O}(a^2)$ effects in the difference between the two masses at the
heaviest quark mass.
\begin{table}[ht]
  \caption{Nucleon masses as measured with Asqtad and domain-wall
  fermions for the valence quarks.}
  \label{tab:nucl-mass}
  \begin{tabular}{@{}*{4}{l}}
    \hline $\Omega$ & $(am)_q^{\text{Asqtad,sea}}$ &
    $(am)_N^{\text{Asqtad}}$ & $(am)_N^{\text{DWF}}$ \\ \hline
    $20^3\times 32$ & $0.050$ & $1.057(5)$ & $1.029(9)$ \\
    & $0.030/0.050$ & $0.930(3)$ & $0.941(10)$ \\
    & $0.010/0.050$ & $0.779(6)$ & $0.756(21)$ \\
    $28^3\times 32$ & $0.010/0.050$ & & $0.763(12)$ \\ \hline
  \end{tabular}
\end{table}

\section{HADRONIC STRUCTURE WITH LIGHT QUARKS}
\label{sec:hadr-struct-with}
Within the framework discussed in the previous sections, we compute
different observables of the nucleon structure for different quark
masses. The results are compared to previous results from full QCD
calculations with heavy quark masses, see
References~\cite{Dolgov:2002zm} for their calculation.

\subsection{Form factor and transverse size of the proton}
\label{sec:form-fact-transv}
First, we consider the electromagnetic form factor $F_1(-t)$ of the
nucleon. Phenomenologically, it follows a dipole-shaped behavior. Its
derivative at the origin specifies the root-mean squared radius,
$r_{\text{MS}}$, of the transverse charge distribution in the
infinite-momentum frame and thus characterizes the size of of the
nucleon. In a hypothetical world with heavy pions, the size of the
nucleon should be smaller due to the absence of a significant pion
cloud.

\begin{figure}[ht]
  \caption{The proton form factor, $F_1(-t)$, as a function of the
    virtuality, $t=q^2$.}
  \label{fig:formfac-F1}
  \vskip 7pt
  \includegraphics[scale=0.25,clip=true]{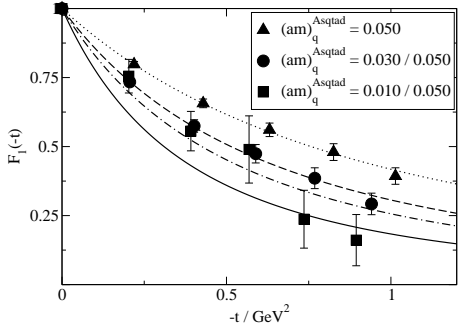}
\end{figure}

This qualitative behavior is well reproduced by our lattice data.
Figure~\ref{fig:formfac-F1} shows our first results for the proton
form factor, $F_1(-t)$. The triangles denote the heaviest quark mass
in Table~\ref{tab:tune-quarks}, the circles the intermediate quark
mass of $(am)_q^{\text{Asqtad,sea}}=0.030/0.050$ and the boxes the
lightest quark mass, all at the volume of $\Omega=20^3\times 32$. The
dotted, the dashed, and the dash-dotted curves show the best dipole
fit to these data points. The solid line shows the best dipole fit to
the experimental data.
\begin{figure}[ht]
  \caption{Transverse root-mean squared radius $r_{\text{MS}}$ of the
    nucleon as a function of the pion mass squared.}
  \label{fig:rms}
  \vskip 7pt
  \includegraphics[scale=0.25,clip=true]{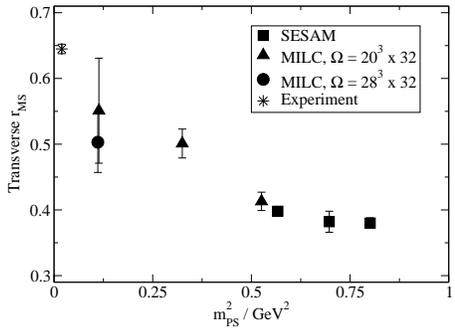}
\end{figure}

Figure~\ref{fig:rms} shows the transverse mean square radius,
$r_{\text{MS}}$. The boxed data points represent the old SESAM
data~\cite{Dolgov:2002zm}, obtained in a full QCD calculation with
Wilson quarks. The triangular data points show the hybrid calculation
with the data points from Table~\ref{tab:tune-quarks} with the volume
$\Omega=20^3\times 32$. Finally, the circle represents the volume
$\Omega=28^3\times 32$ in Table~\ref{tab:tune-quarks}.  As the quark
mass decreases, the transverse size of the nucleon increases and
approaches the experimental value. Note that we find no evidence of
substantial finite-size corrections since there is no statistically
significant difference between the data points at different volumes.

\subsection{Axial charge}
\label{sec:axial-charge}
The axial charge, $g_A$, is given by the forward value of the
generalized parton distribution $\tilde{A}_1^{\text{u-d}}$ (see
Reference~\cite{Hagler:2003jd} for details). This quantity has been
shown to exhibit substantial finite-size
dependence~\cite{Sasaki:2003jh} and therefore provides a good
benchmark of the importance of the physical box size as we approach
the chiral limit.

\begin{figure}[htb]
  \caption{Nucleon axial charge, $g_A$, as a function of the pion mass
    squared.}
  \label{fig:gA}
  \vskip 7pt
  \includegraphics[scale=0.25,clip=true]{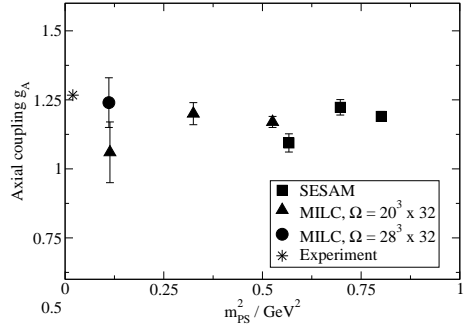}
\end{figure}

Figure~\ref{fig:gA} shows a central result of this work, the nucleon
axial charge as a function of the pion mass squared. It has been
renormalized using the five-dimensional conserved current. The symbols
are the same as in Figure~\ref{fig:rms}.

We observe that the SESAM results in a $(1.6\,\mbox{fm})^3$ box first
increase with decreasing pion mass and then fall off when the pion
cloud becomes too large to fit in the box. In the larger
$(2.6\,\mbox{fm})^3$ MILC box (on a $20^3$ lattice), the value at
$m_\pi\simeq 700$ MeV increases back to its physical value again and
then falls off at $m_\pi\simeq 300$ MeV when the light pion no longer
fits in the box. Increasing the box to $(3.5\,\mbox{fm})^3$ ($28^3$
lattice) allows $g_A$ to increase to a physical value consistent with
experiment. Thus, the locus of points in the largest box at each mass
extrapolate smoothly to the experimental result.

To summarize, we give the resulting values of $r_{\text{MS}}$ and
$g_A$ in Table~\ref{tab:hadstruc}.
\begin{table}[ht]
  \caption{Summary of the hadron structure results. The table shows
  the volume, the Asqtad sea quark mass from
  Table~\ref{tab:tune-quarks}, and the resulting values of
  $r_{\text{MS}}$ and $g_A$.}
  \label{tab:hadstruc}
  \begin{tabular}{@{}*{4}{l}}
    \hline $\Omega$ & $(am)_q^{\text{Asqtad,sea}}$ & $r_{\text{MS}}$ &
    $g_A$ \\ \hline
    $20^3\times 32$ & $0.050$ & $0.4113(14)$ & $1.17(2)$ \\
    & $0.030/0.050$ & $0.501(22)$ & $1.20(4)$ \\
    & $0.010/0.050$ & $0.551(80)$ & $1.06(11)$ \\
    $28^3\times 32$ & $0.010/0.050$ & $0.503(46)$ & $1.24(9)$ \\
    \hline
  \end{tabular}
\end{table}

\section{SUMMARY AND OUTLOOK}
\label{sec:summary-outlook}
In this talk we have used a tuned hybrid calculation of Asqtad sea
quarks and domain-wall valence quarks. We have determined an optimal
value for $L_5$ in the domain-wall action and tuned the bare quark
masses such that the pion masses of an Asqtad-valence and the
domain-wall valence calculation agree. If the continuum limit exists,
this calculation provides a valid scheme to compute all hadronic
observables.

We have computed the $F_1(-t)$ form factor of the proton and obtained
the $r_{\text{MS}}$ radius from it. The results are in qualitative
agreement with the picture of the nucleon becoming larger as the pion
cloud becomes more prominent with decreasing quark mass.

We have also applied our scheme to the case of the nucleon axial
coupling, $g_A$. Whereas calculations in a fixed volume underestimate
$g_A$ when the pion becomes too light to fit in the box, our sequence
of calculations in three volumes produce a locus of points that
smoothly extrapolate to the experimental result.

Encouraged by the successes reported here, we are continuing the
calculation of hadronic observables using improved staggered sea
quarks and domain-wall valence quarks.

\end{document}